\documentclass[aps,prl,twocolumn,showpacs,groupedaddress]{revtex4}
\usepackage{graphicx}

\begin{document}               

\def\be{\begin{equation}}
\def\ee{\end{equation}}
\def\bd{\begin{displaymath}}
\def\ed{\end{displaymath}}
\def\ba{\begin{eqnarray}}
\def\ea{\end{eqnarray}}
\def\lr{\leftrightarrow}
\def\s{\protect{f_0}}
\def\C{{\rm C}}
\def\D{{\rm D}}
\def\L{{\rm L}}
\def\HI{{\rm H}_{\rm I}}

\title{Charmonium Cross Sections and the QGP}
\author{T.Barnes}
\email{tbarnes@utk.edu}
\affiliation{Physics Division, 
Oak Ridge National Laboratory, Oak Ridge, TN 37831, USA\\
Department of Physics and Astronomy, 
University of Tennessee,\\
Knoxville, TN
37996, USA 
}
                                                                                
%
\begin{abstract}
In this short review
we summarize experimental information and theoretical
results for the low-energy 
dissociation cross sections of charmonia by light hadrons.
These cross sections are required for the simulation of 
charmonium absorption through collisions with comovers
in heavy ion collisions,
which competes with quark-gluon plasma production
as a charmonium-suppression mechanism.
If the cross sections are sufficiently large 
these dissociation reactions may be misinterpreted as 
an effect of quark-gluon plasma production.
Theoretical predictions for these RHIC-related
processes have used various methods, including 
a color-dipole scattering model, meson exchange models,
constituent interchange models and QCD sum rules.
As the results have been largely unconstrained by experiment,
some of the predictions differ by orders of magnitude, notably in the 
near-threshold regime that is most relevant to QGP searches.
\end{abstract}


\maketitle

\section{Introduction}
\label{intro}

Many unusual subjects have been studied in the name of QCD.
One of the more unusual, which has arisen in the field of heavy ion
collisions, is the size of
cross sections of charmonia on light hadrons.
This has attracted attention because of the proposal by
Matsui and Satz \cite{Matsui:1986dk} that suppression of $J/\psi$
production could be used as
a signature for the formation of a quark-gluon plasma.
                                                                                
\begin{figure}[b]
\includegraphics[width=0.3\textwidth]{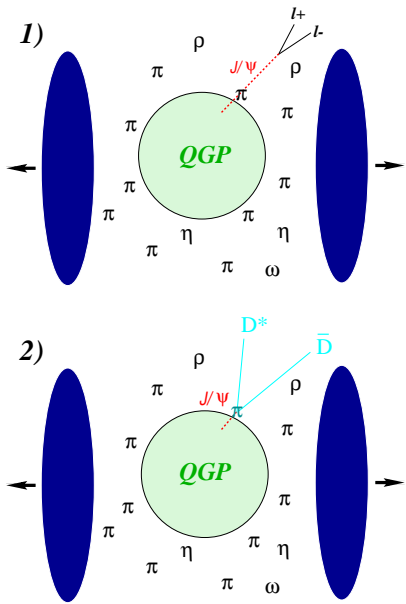}
\vskip -0.3cm
\caption{We wish to distinguish between
1) weak and
2) strong
$J/\psi$ absorption by light comovers.}
\label{fig:fig1}
\end{figure}

This
suggestion,
like many signatures proposed for the quark-gluon plasma,
is perhaps excessively intuitive. The idea is that a QGP will
screen the linear confining interaction between quarks,
so that a $c\bar c$ pair produced within a QGP will be
less likely to form a bound $c\bar c$ charmonium resonance,
as in Fig.1,
but instead will more likely separate to form open-charm mesons.

Even if this simple picture of $c\bar c$ production in a QGP
is qualitatively correct, it can only be confirmed easily if
the competing direct charm production and scattering by the
initial relativistic nucleons is understood 
\cite{Hufner:1998hf,Hufner:1999ak,Wong:wm} and
if there is little subsequent dissociation of the charmonia
by the many other
``comoving" light hadrons produced in such a collision.
To summarize the last point,
if charmonium + light hadron ``comover"
dissociation cross sections
are small
(case 1, top of Fig.1)
and the
background of direct
charm production
from the initial nucleons is understood,
one may have a useful signature of QGP formation, but
if the comover dissociation
cross sections are large
(case 2, bottom of Fig.1)
one must distinguish
a QGP-reduced charmonium production amplitude from
subsequent dissociative scattering, and the interpretation of this
signal will therefore be ambiguous.
                                                                                
Thus it is of great relevance to the interpretation of
RHIC physics to establish the
approximate size of these low-energy
$c\bar c$ + light hadron cross sections.
                                                                                
\section
{Experiment, or what passes for it}
Unfortunately we have no charmonium beams or targets, so the
experimental cross sections
must be inferred
indirectly
and are poorly known.
The earliest estimates of lower energy charmonium
hadronic cross sections
came from $J/\psi$ photoproduction experiments in the
mid 1970s, which were interpreted in terms of a  $J/\psi$+N
cross section given additional theoretical assumptions.
Early Fermilab and SLAC photoproduction
experiments gave
rough estimates of $\sim 1$ mb for
$\sigma_{J/\psi+{\rm N}}$,
assuming vector dominance, for
photon energies from $E_\gamma \approx 13$ to $200$ GeV 
\cite{Knapp:nf,Camerini:1975cy}.
A subsequent SLAC photoproduction
experiment in 1977 used the $A$ dependence of
$J/\psi$ absorption to estimate a rather larger cross section
of
$\sigma_{J/\psi+{\rm N}} = 3.5 (0.8)$ mb
at $E_\gamma \approx 17$ GeV
($\sqrt{s} \approx 6$ GeV)
\cite{Anderson:1976hi}.
The vector dominance hypothesis may have lead to an underestimate of
the cross section in the earlier references \cite{Hufner:1997jg}.
                                                                                
In heavy ion collisions these cross sections may be estimated from
the ratio of lepton pairs produced in the
$J/\psi$ peak to ``background" Drell-Yan pairs nearby in energy.
Since the  $J/\psi$ must reach the exterior of the nuclear target
to decay into a sharp mass
peak, this ratio gives us an estimate of the
absorption cross section
through a classical survival probability formula,
\begin{displaymath}
\sigma(J/\psi\to \mu^+\mu^-) / \sigma({\rm Drell-Yan}\ \mu^+\mu^-)
\end{displaymath}
\begin{equation}
=
\exp(-\rho\; \sigma_{J/\psi + {\rm N}}^{abs}\; {\rm L} )
\end{equation}
where $\rho$ is the mean nucleon density and L is the estimated
mean path length in
the experimental nuclear system.
A ``naive" interpretation of the
$J/\psi$ production data from collisions of various nuclear species
using this formula
gives
$\sigma_{J/\psi + {\rm N}}^{abs}\approx 6$ mb at
$\sqrt{s}\approx 10$ GeV
\cite{Capella:1996va}, with a numerically similar result for the $\psi'$.
                                                                                
\begin{figure}
\includegraphics[width=0.4\textwidth]{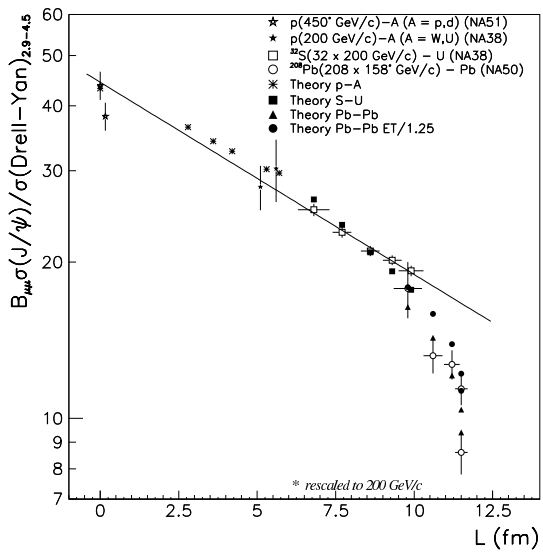}
\vskip -0.3cm
\caption{A fit of Eq.(1) to experimental $J/\psi$ production versus
path length \cite{Capella:1996va}; the line corresponds to 6.2 mb.}
\label{fig:fig2}
\end{figure}

Of course one may raise many questions about the validity of
this simple estimate, including
the importance of shadowing in Drell-Yan, the neglect of
$J/\psi$ scattering by other light hadrons formed in the collision
(for example $\pi$ and $\rho$), and the assumption of
a single, constant $J/\psi+$N cross section in all circumstances.

In addition to these more or less direct measurements there
have been several attempts to infer charmonium cross sections
from related processes, given additional theoretical assumptions.
A series of theoretical papers
\cite{Sorge:1997bg,Chen:1997zz,Fujii:1999xn,Barnes:2003kp}
has estimated the rather weak closed-flavor 
cross sections such as $J/\psi + \pi \to \psi' + \pi$ near threshold
from the observed dipion decays (here $\psi' \to \pi \pi J/\psi$). 
Redlich {\it et al} \cite{Redlich:2000cb} have used charm photoproduction
data combined with a vector dominance model to estimate the 
$J/\psi {\rm N}$ cross section, and find quite small values near
threshold.  H\"ufner {\it et al.} \cite{Hufner:2000jb}
again argue however that vector dominance is not justified for 
this process, so that the $J/\psi + {\rm N}$ cross sections
estimated in this manner are inaccurate.

Recently, concerns have been expressed that the
$J/\psi$ and
$\psi'$
wavefunctions have not had sufficient time to form within the
nucleus in these collisions, so experiment may instead be measuring
the cross section for a small initial $c\bar c$ ``premeson" on a nucleon.
One can increase the time spent in the interior of the nuclear system
by selecting small
and even negative $x_F$ events, as has been done by E866 at Fermilab.
As discussed by
He, H\"ufner and Kopeliovich
\cite{Hufner:1995qe,He:1999aj}, this leads one to infer
$\sigma_{J/\psi + {\rm N}}^{abs} = 2.8 (0.3)$ mb
and
$\sigma_{\psi' + {\rm N}}^{abs} = 10.5 (3.6)$ mb respectively, also at
$\sqrt{s}\approx 10$ GeV.
This is rather more satisfying to people who have an
intuitive notion that the larger $\psi'$ should have a
larger reaction cross section. Actually the connection between
cross section and physical extent is less direct (compare
KN and $\bar {\rm K}$N),
and in any case the relative proximity of
inelastic thresholds alone would suggest a
larger $\psi'$ cross section. These experiments also indicate a
preference for dissociation over elastic cross sections
in this energy region by roughly a factor of 30 \cite{Hufner:1995qe,He:1999aj}.
                                                                                
\section{
Theory: Introduction}
                                                                                
To quote B.M\"uller in 1999,
``...the state of the theory of interactions
between $J/\psi$ and light hadrons is embarrassing [...].
Only three serious calculations exist (after more than 10 years of
intense discussions about this issue!) and their results differ
by at least two orders of magnitude in the relevant energy range
[...]. There is a lot to do for those who would like to make a
serious contribution to an important topic." \cite{Muller:1999ys}.
                                                                                
The theoretical situation has improved considerably in recent
years,
at least in terms of the number of calculations if not in
the understanding of the scattering mechanism.
A list of $c\bar c$ + light hadron cross section calculations
is given in Table I.

\subsection{Color Dipole Model}

The most cited work, {\it albeit} furthest in its predictions from
a low-energy
``theoretical mean", is the color-dipole scattering 
calculation of Kharzeev and Satz
\cite{Kharzeev:1994pz}.
This reference is basically a restatement of the
color-dipole model
developed in the late 1970s by Peskin and Bhanot 
\cite{Peskin:1979va,Bhanot:1979vb,Kaidalov:hd}
for scattering of light hadrons by
Coulombic bound states of very massive quarks.
According to
Peskin, the criterion for validity of this approach is
``...not met even for the $b\bar b$
system."
\cite{Pes00}, so there may be large systematic errors at the
$c\bar c$ mass scale.
This approach certainly makes marginal approximations
for charmonium, such as the use of Coulombic wavefunctions
(which are far from accurate for $c\bar c$) and the
introduction of a
$Q\bar Q$ binding
energy (which is hard to interpret for charmonium, and
is
taken to be $2\, M_\D  - M_{J/\psi}$ by Kharzeev and Satz).
These color-dipole scattering
formulas also implicitly assume that charmonia are small
relative to
the natural QCD length scale.
Since potential
models actually find rms $c\bar c$ separations
of about 0.4~fm for the $J/\psi$, 0.6~fm for the $\chi_c$ states and
0.8~fm
for the $\psi'$ \cite{Buchmuller:1980su}, this is also a dubious assumption.
                                                                                
\begin{table}
\begin{tabular}{|l||l|c|}
\hline
Method & Init. State & Ref. \cr
\hline
\hline
color dipole  &
$J/\psi + $N  &
\cite{Kharzeev:1994pz}
\cr
            &
$J/\psi + (\pi, $N)  &
\cite{Arleo:2001mp}
\cr
            &
$(J/\psi, \psi') + $N  &
\cite{Oh:2001rm,Hufner:2000jb}
\cr
\hline
meson ex.   &
$J/\psi + \pi $ &
\cite{Blaschke:2000zm}  \cr
                 &
$J/\psi + \pi, \rho $ &
\cite{Matinian:1998cb,Lin:1999ad,Oh:2000qr,Oh:2002vg,Haglin:2003fh}  \cr
                 &
$J/\psi + $N  &
\cite{Sibirtsev:2000aw,Liu:2001yx}  \cr
                 &
$J/\psi + \pi, K, \rho, $N  &
\cite{Haglin:1999xs,Haglin:2000ar}  \cr
\hline
constit. int.  &
$J/\psi + \pi$ &
\cite{Blaschke:2000zm,Martins:1994hd} \cr
  &
$(J/\psi, \psi') + (\pi, \rho$) &
\cite{Wong:1999zb} \cr
  &
$(J/\psi, \psi', \chi_J) + (\pi, \rho, {\rm K})$ &
\cite{Wong:2001td} \cr
  &
$(J/\psi, \psi') + (\pi, \rho) \; ;\ \chi_J + \pi $&
\cite{Barnes:2003dg} \cr
  &
$J/\psi  + $N &
\cite{Black:2002} \cr
  &
$(J/\psi, \psi' ) + ( \pi, $N )&
\cite{Martins:1995nb} \cr
\hline
QCD sum rules &
$J/\psi + \pi $ &
\cite{Navarra:2001jy,Duraes:2002ux}  \cr
\hline
\end{tabular}
\caption{A summary of near-threshold
$c\bar c$ + light hadron cross section calculations.}
\end{table}
                                                                                
Although this approach has problems with justification
for $c\bar c$, the predictions are
nonetheless interesting as estimates of the scale of these cross
sections assuming a color-dipole scattering mechanism,
and the approach could presumably be extended to lower energies
by generalizing the
wavefunctions and interaction.
The formula for the $J/\psi+$N cross section
quoted by Kharzeev and Satz \cite{Kharzeev:1994pz} is
\begin{equation}
\sigma_{J/\psi+{\rm N}} = 2.5\hbox{ mb}\ \bigg(1 - {\lambda_0 \over \lambda}
\bigg)^{6.5} 
\end{equation}
where $\lambda = (s-M_{J/\psi}^2 - M_{\rm N}^2) / 2\, M_{J/\psi}$,
the constant
$\lambda_0$ is defined to be
$ `` \simeq  M_{\rm N} + \epsilon_0 "$ according to the text following
Eq.(24) of \cite{Kharzeev:1994pz} (we assume the equality),
and the ``binding energy" $\epsilon_0$ is set to $2\, M_\D  - M_{J/\psi}$.
The result
is shown in Fig.3, together with the
single lower-energy SLAC experimental point
\cite{Anderson:1976hi}.
                                                                                
\begin{figure}
\includegraphics[width=0.4\textwidth]{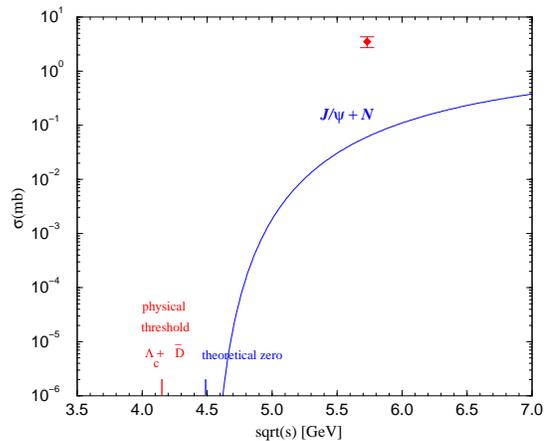}
\vskip -0.3cm
\caption{The Kharzeev-Satz $J/\psi+$N total cross section
and the 1977 SLAC result \cite{Anderson:1976hi}.}
\label{fig:fig3}
\end{figure}
                                                                                
Evidently the Kharzeev-Satz cross section is smaller than this SLAC
point (which was an inferred cross section and certainly needs
confirmation) by about two orders
of magnitude, and falls precipitously as $\sqrt{s}$ is decreased.
Their calculation
actually has no direct information about physical thresholds,
so Kharzeev~{\it et}~{\it al.} typically leave their curves ``dangling" just
below
$\sqrt{s} = 5$~GeV. (See Fig.2 of \cite{Kharzeev:1994pz} for example.)
If we plot their formula Eq.(2) for
$J/\psi+$N at low energy (Fig.3), we find the unphysical prediction of a
zero cross section near 4.5~GeV, whereas the physical
$\Lambda_c + \bar \D$ threshold is at 4.15~GeV. Obviously 
this calculation is
inapplicable at low energies, which is unfortunate
because this is the regime of greatest
interest for QGP searches.

Similar cross section calculations have since been reported by
Arleo {\it et al.} \cite{Arleo:2001mp} and
Oh {\it et al.} \cite{Oh:2001rm} (also using Coulomb wavefunctions,
but incorporating physical kinematical constraints)
and by 
H\"ufner {\it et al.} \cite{Hufner:2000jb},
using a similar dipole scattering model
with more realistic $c\bar c$
wavefunctions. 
Oh {\it et al.} \cite{Oh:2001rm} note that this model
predicts an extremely large cross
section ratio of
$\sigma_{\psi' + p} / \sigma_{J/\psi + p}\sim 2000$-$5000$
at 4.2~GeV. 
H\"ufner {\it et al.} find 
$J/\psi + p$ and $\psi' + p$ 
cross sections of 
${\approx }$ 
3~mb and 10~mb 
respectively
at $\sqrt{s} = 10$~GeV (see their Fig.10),
and do not consider use of this approach 
justified near threshold.

\subsection{Meson Exchange Models}
                                                                                
Several calculations of charmonium + light hadron
cross sections
have been reported
assuming $t$-channel charmed meson exchange.
Of course this picture is also problematic, since the
range of the exchanged charmed meson would be only about $1/M_\D 
\approx 0.1$ fm, and the assumption of nonoverlapping hadrons 
at this separation
is clearly invalid.
(This is the Isgur-Maltman \cite{Maltman:st} argument as to
why vector meson exchange is
unjustified as the source of the short-ranged NN core interaction.)
Nonetheless it is again interesting to see what scale of
cross section is predicted by this type of model,
since it might at least incorporate the correct scales and
degrees of freedom, and it assumes a different scattering mechanism
from the color-dipole model advocated by Kharzeev and Satz.

The first such meson exchange calculation, due to Matinian and M\"uller
\cite{Matinian:1998cb}, considered $t$-channel D exchange as the mechanism
for the reactions
$J/\psi+\pi\to \D^*\bar \D + h.c.$ and $J/\psi+\rho\to \D\bar \D$;
their results are shown in Fig.4. Note that 500~MeV
above threshold these cross sections lie in the 0.5 to 1 mb range.
A subsequent calculation by Haglin \cite{Haglin:1999xs},
who assumed an SU(4) invariant gauge field effective meson lagrangian,
found few-mb scale results for these cross sections near threshold.

Similar meson exchange dissociation 
cross sections calculations 
for $J/\psi$+N were reported
by Haglin \cite{Haglin:1999xs},
Sibirtsev, Tsushima and Thomas
\cite{Sibirtsev:2000aw} and
Liu, Ko and Lin
\cite{Liu:2001yx}. 
Haglin found
a peak of about 7 mb near $\sqrt{s}=4.3$ GeV,
whereas
Sibirtsev {\it et al.} found a peak cross section of about 1 mb near
$\sqrt{s}=4.6$ GeV (Fig.5). Liu {\it et al.} found a scale similar
to Sibirtsev {\it et al.} near threshold, but concluded that production 
of charmed meson pairs dominated at higher energies.
Since Kharzeev and Satz found that the 
color-dipole scattering model predicts negligible
$J/\psi+$N cross sections at low energies (Fig.3),
evidently flavor-exchange processes are predicted to be 
dominant near threshold.

\begin{figure}
\includegraphics[width=0.4\textwidth]{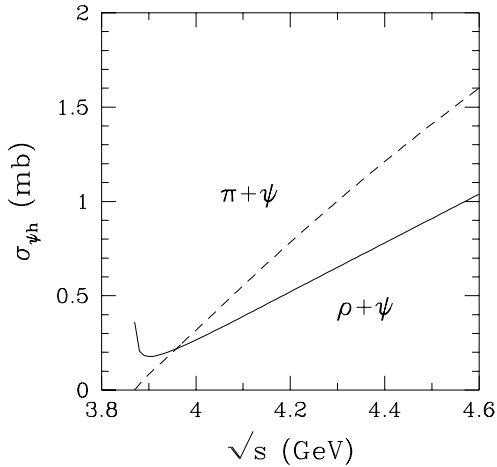}
\vskip -0.3cm
\caption{The Matinian-M\"uller $t$-channel meson exchange results for
$J/\psi+\pi$
and
$J/\psi+\rho$ inelastic cross sections
\cite{Matinian:1998cb}.}
\label{fig:fig4}
\end{figure}

\begin{figure}
\includegraphics[width=0.4\textwidth]{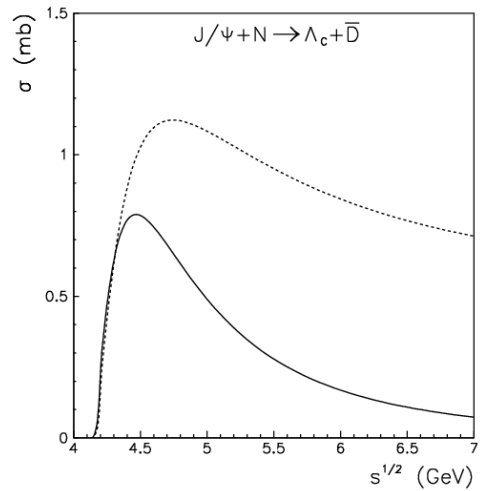}
\vskip -0.3cm
\caption{The $t$-channel meson exchange cross sections
for $J/\psi+{\rm N} \to \Lambda_c+\bar \D$
found by Sibirtsev, Tsushima and Thomas
\cite{Sibirtsev:2000aw}.
The smaller contribution is from $\D$ exchange
and the larger is from (non-interfering) $\D^*$ exchange.}
\label{fig:fig5}
\end{figure}

Subsequent meson-exchange studies 
by Lin and Ko \cite{Lin:1999ad}, Haglin and Gale \cite{Haglin:2000ar}
and Oh {\it et al.} \cite{Oh:2000qr,Oh:2002vg} 
found that the assumption
of pointlike mesons in the effective lagrangian leads to unrealistically
large cross sections even at moderate energies, so that hadronic form factors
must be incorporated in the calculations. With plausible but rather arbitrary 
form factors these cross sections are greatly reduced, so that they are again 
found to be typically of few-mb scale (see Fig.6 for example). 

\begin{figure}
\includegraphics[width=0.4\textwidth]{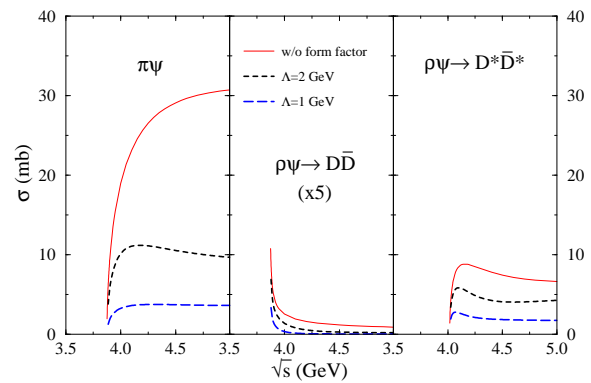}
\vskip -0.3cm
\caption{A strong suppression of dissociation cross sections
is found on incorporating hadronic form factors in meson 
exchange models.
This example is Fig.4 of Lin and Ko \cite{Lin:1999ad}. 
}
\label{fig:fig6}
\end{figure}

\eject
In summary, the current, rather unsatisfying state of affairs in
meson exchange models of charmonium dissociation 
is that mb-scale cross sections are anticipated 
near threshold, but their precise values depend on poorly known 
hadronic form factors, as well as on effective meson lagrangians 
of unknown accuracy.
Future studies that can put the effective lagrangians and form factors 
on a more sound theoretical basis would be an important contribution 
to this approach. One interesting possibility, 
which has recently been investigated by
Navarra~{\it et}~{\it al.} and
Matheus {\it et al.},
is to derive the hadronic form factors from QCD sum rules 
\cite{Navarra:2001ju,Matheus:2002nq}.
An especially attractive possibility 
explored very recently by Deandrea, Nardulli and Polosa
\cite{Deandrea:2003pv} 
is that the hadronic 
form factors may be evaluated explicitly 
in terms of quark model wavefunctions.

\vskip -0.3cm
\subsection{Constituent Interchange Model}
                                                                                
In this approach one uses explicit 
nonrelativistic
quark model wavefunctions for the external hadrons 
and calculates the cross section
assuming a constituent interchange scattering
mechanism, driven by the Born-order matrix element of
the standard quark model Hamiltonian.
Constituent interchange
$c\bar c$ + $q\bar q$ 
cross sections have
their strongest support just a few hundred MeV
in $\sqrt{s}$ above threshold,
since the overlap integrals are damped by the external
meson wavefunctions at higher momenta.

This technique, which has no free parameters
once quark model wavefunctions
and the interquark Hamiltonian are specified,
has been shown to compare reasonably well with
experimental
low-energy hadron-hadron scattering data near threshold
for a wide range of annihilation-free reactions.
There are four quark interchange diagrams in meson-meson scattering
(Fig.7),
each of which has an associated overlap integral of the 
external meson wavefunctions convolved with the interquark
Hamiltonian. Constituent interchange is forced at Born-order
because $\HI\propto \lambda^a \cdot \lambda^a $ changes each
initial color-singlet $q\bar q$ meson into a color octet, which has overlap
with color-singlet final state mesons only after quark line interchange.
Feynman rules for these diagrams were given by Barnes and Swanson
\cite{Barnes:1991em}.
                                                                                
\begin{figure}[h]
\includegraphics[width=0.35\textwidth]{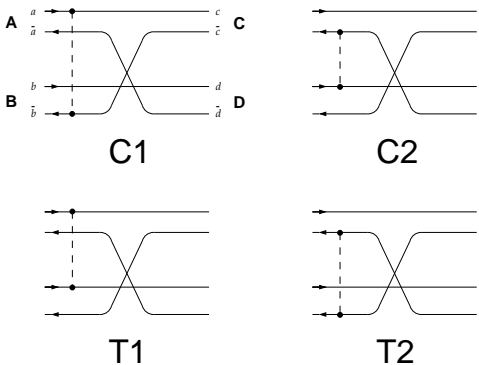}
\vskip -0.3cm
\caption{The four constituent interchange scattering diagrams evaluated in 
$c\bar c$ + $q\bar q$ cross section calculations (prior formalism).
The ``exchange" is the full
quark-quark interaction Hamiltonian $\HI$.}
\label{fig:fig7}
\end{figure}

The first charmonium cross section calculation using this approach
was due to
Martins, Blaschke and Quack
\cite{Martins:1994hd},
who considered the reactions
$J/\psi+\pi \to \D^* \bar \D +h.c.$
and
$\D^* \bar \D^* $ (The amplitude for $J/\psi+\pi\to \D\bar \D$ is zero in
the nonrelativistic quark model
without spin-orbit forces, 
and has been found to be quite weak in a relativized
calculation \cite{Blaschke:2000zm}.)
Martins {\it et al.} found that these exclusive final states have
numerically rather similar cross sections
(except for their different thresholds),
and give a
maximum total cross section of about 7 mb at
$\sqrt{s}\approx 4.1$ GeV.
\begin{figure}[ht]
\includegraphics[width=0.32\textwidth]{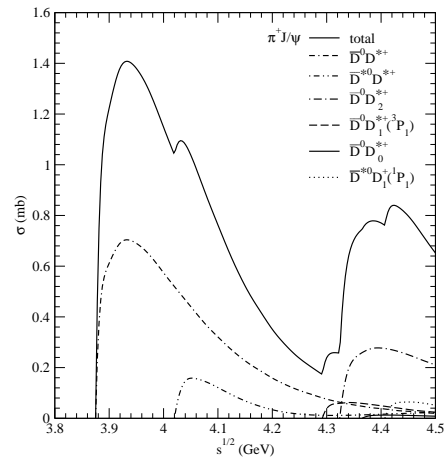}
\vskip -0.3cm
\caption{Constituent interchange model predictions for 
$J/\psi + \pi $ exclusive and total
cross sections
\cite{Barnes:2003dg}. 
}
\label{fig:fig8}
\end{figure}
\begin{figure}[ht]
\includegraphics[width=0.32\textwidth]{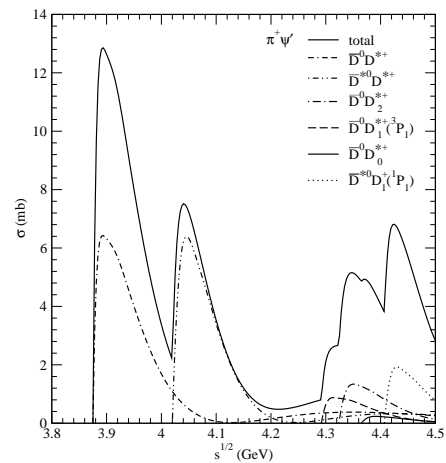}
\vskip -0.3cm
\caption{$\psi' + \pi $ 
cross sections
\cite{Barnes:2003dg}. Note the change of scale relative to Fig.8.
}
\label{fig:fig9}
\end{figure}
\begin{figure}[ht]
\includegraphics[width=0.32\textwidth]{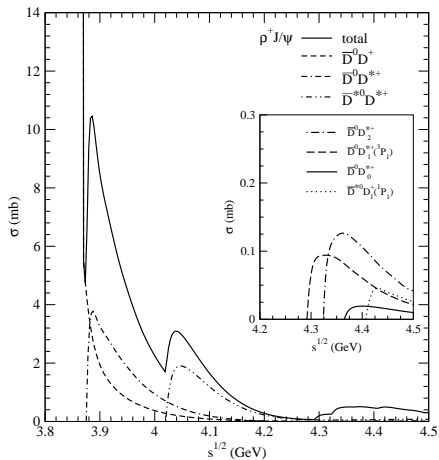}
\vskip -0.5cm
\caption{$J/\psi + \rho $
cross sections
\cite{Barnes:2003dg}.
}
\label{fig:bnew_fig10}
\end{figure}
Wong {\it et al.} have 
carried out similar constituent interchange cross section
calculations \cite{Wong:1999zb,Wong:2001td,Barnes:2003dg}, using
numerically determined Coulomb plus linear plus smeared hyperfine
quark potential model
wavefunctions, with parameters fitted to the 
$q\bar q$ and $c\bar c$ meson spectra.
Figs.8-10 show some recent results from Ref.\cite{Barnes:2003dg}.
The $J/\psi + \pi $ cross section is somewhat smaller 
than was found by Blaschke {\it et al.}, and peaks
at about 1.4~mb just above
3.9~GeV. 

The difference lies mainly in the treatment
of the confining interaction; 
Blaschke {\it et al.} treated confinement
as a color-independent Gaussian potential between
quark and antiquark (hence they include only diagrams C1 and C2),
whereas Wong {\it et al.} used the conventional
$\lambda^a \cdot \lambda^a$ form
between all pairs of constituents.
This leads to
destructive interference between the C and T diagrams due to 
opposite color factors, and hence to
a much reduced total cross section. 
Ref.\cite{Barnes:2003dg} also considers 
$\psi' + \pi $ scattering; the rather
large cross section found for this process 
is shown in Fig.9.
As a final example, the constituent-interchange 
$J/\psi + \rho $ cross section from Ref.\cite{Barnes:2003dg}
is shown in Fig.10; this diverges at
threshold for the simple reason that the D$\bar {\rm D}$ 
channel is exothermic.

\eject       
It is interesting that the simple
two-parameter
function 
\begin{equation}
\sigma(s) = \sigma_{max} \;
(\epsilon / \epsilon_{max} )^p \;
e^{p(1-\epsilon / \epsilon_{max} )}  \ ,
\end{equation}
often provides an accurate parametrization of the constituent
interchange cross sections.
In this formula 
$\epsilon =\sqrt{s} - M_\C - M_\D$ is the $c.m.$ energy above threshold 
and $\sigma_{max}$ is the cross section maximum, at
$\epsilon_{max}$.
The threshold exponent $p$ is determined by the angular quantum numbers of
the hadrons, and is
$\pm 1/2 + \L_{min}^{\C\D}$ (for endothermic/exothermic),
where $\L_{min}^{\C\D}$ is the lowest angular momentum
allowed for the final meson pair $\C\D$. 

\vskip 0.2cm
The much more complicated problem of charmonium-nucleon scattering
has also been investigated in the constituent interchange
model. 
$J/\psi + $N and $\psi' + $N cross sections
were derived by Martins \cite{Martins:1995nb}
using a simplified quark+diquark model of the nucleon;
this approach gave several-mb peak cross sections 
not far above threshold.
Black \cite{Black:2002} 
has carried out the full $J/\psi + $p 
constituent interchange calculation 
for several final states, assuming hyperfine, Coulomb and linear 
interactions. He finds that the final state 
$\bar {\rm D}^o \Lambda_c^+$ is dominant, with a surprisingly large 
({\it ca.} 12~mb) peak cross section just above threshold.

\subsection{QCD Sum Rules}

The application of QCD sum rules to the determination of charmonium 
dissociation cross sections, due to 
Navarra {\it et al.}
\cite{Navarra:2001jy}
and to Dur\~aes {\it et al.}
\cite{Duraes:2002ux},
is a relatively recent development.
This method relates the scattering amplitude to a sum of 
operator vacuum expected values, and gives a model independent 
result to the extent that these expected values are known experimentally
and the set of operators chosen does indeed dominate the scattering amplitude
over the chosen kinematic regime. 
There are also systematic uncertainties in the approach
due to the details of a Borel summation and treatment of a continuum 
contribution to the amplitudes.

\begin{figure}
\includegraphics[width=0.35\textwidth]{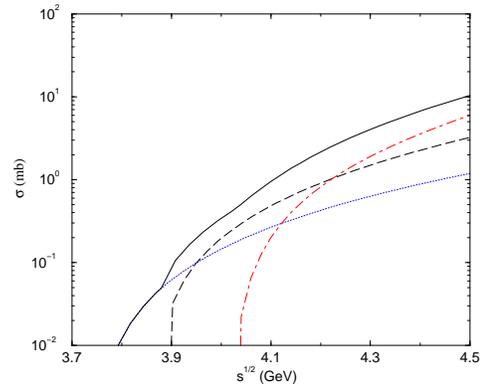}
\vskip -0.5cm
\caption{QCD sum rule results found by Navarra {\it et al.} 
\cite{Navarra:2001jy} for the cross sections 
$J/\psi + \pi \to \D\bar\D,\ \D^*\bar\D + h.c.$ and $\D^*\bar\D^*$
(distinguished by the thresholds) and their sum (solid).
}
\label{fig:bnew_fig11}
\end{figure}

The studies published to date 
are a calculation of the
$J/\psi$  
dissociation reactions
$J/\psi + \pi \to \D\bar\D, \D^*\bar\D + h.c.$ and $\D^*\bar\D^*$
\cite{Navarra:2001jy},
followed by a more detailed investigation of the same processes
\cite{Duraes:2002ux}.
The sum rule results (Fig.11) appear to confirm the mb-scale of 
near-threshold dissociation cross sections, in 
qualitative agreement 
with both meson exchange and constituent interchange models.

\section{Conclusions}

We have reviewed theoretical predictions and related 
experimental results for 
the low-energy dissociation cross sections of charmonia 
on light hadrons, which are of great importance for QGP searches 
in heavy ion collisions.
Four theoretical approaches have recently been applied to
this problem, which are a color-dipole scattering model, 
meson exchange, constituent interchange, and QCD sum rules.
Near threshold the color-dipole model predicts very small cross sections,
whereas meson exchange, constituent interchange and QCD sum rules
all predict mb-scale cross sections. At present there are no direct 
measurements of these dissociation 
cross sections near threshold. It would clearly be of great interest
to measure any of these low-energy dissociation cross sections
experimentally, both for the relevance to QGP searches and as a 
valuable test of the theoretical models of hadron scattering that 
have been applied to this problem.

\section{Acknowledgements}
This is an expanded and updated version of a presentation to the 
International Conference on Quark Nuclear Physics
QNP-2002, held at J\"ulich in June 2002. 
It is a pleasure to
acknowledge the kind invitation and support of the
organizers, which made it possible to attend the meeting 
and present a preliminary version of this report.
This research was supported in part by the 
the University of Tennessee,
the U.S. Department of Energy
Division of Nuclear Physics
at ORNL,
managed by UT-Battelle, LLC,
under Contract No. DE-AC05-00OR22725,
Research Corp,
and by the Deutsche Forschungsgemeinschaft DFG under contract 
Bo56/153-1.
We would also like to thank D.B.Blaschke, K.Haglin, J.H\"ufner, 
B.Kopeliovich, M.Peskin,
A.Sibirtsev, S.Sorensen, E.S.Swanson, C.Y.Wong and X.M.Xu 
for useful discussions and information.


\begin{thebibliography}{99}

\bibitem{Matsui:1986dk}
T.~Matsui and H.~Satz,
Phys.\ Lett.\ B {\bf 178}, 416 (1986).

\bibitem{Hufner:1998hf}
J.~H\"ufner and B.~Z.~Kopeliovich,
Phys.\ Lett.\ B {\bf 445}, 223 (1998)
[arXiv:hep-ph/9809300].

\bibitem{Hufner:1999ak}
J.~H\"ufner, Y.~B.~He and B.~Z.~Kopeliovich,
Eur.\ Phys.\ J.\ A {\bf 7} (2000) 239
[arXiv:hep-ph/9908244].

\bibitem{Wong:wm}
C.~Y.~Wong,
Nucl.\ Phys.\ A {\bf 610}, 434C (1996).

\bibitem{Knapp:nf}
B.~Knapp {\it et al.},
Phys.\ Rev.\ Lett.\  {\bf 34}, 1040 (1975).

\bibitem{Camerini:1975cy}
U.~Camerini {\it et al.},
Phys.\ Rev.\ Lett.\  {\bf 35}, 483 (1975).
\bibitem{Anderson:1976hi}
R.~L.~Anderson {\it et al.},
Phys.\ Rev.\ Lett.\  {\bf 38}, 263 (1977).

\bibitem{Hufner:1997jg}
J.~H\"ufner and B.~Z.~Kopeliovich,
Phys.\ Lett.\ B {\bf 426}, 154 (1998)
[arXiv:hep-ph/9712297].

\bibitem{Capella:1996va}
A.~Capella, A.~Kaidalov, A.~Kouider Akil and C.~Gerschel,
Phys.\ Lett.\ B {\bf 393}, 431 (1997)
[arXiv:hep-ph/9607265].

\bibitem{Sorge:1997bg}
H.~Sorge, E.~V.~Shuryak and I.~Zahed,
Phys.\ Rev.\ Lett.\  {\bf 79}, 2775 (1997)
[arXiv:hep-ph/9705329].

\bibitem{Chen:1997zz}
J.~W.~Chen and M.~J.~Savage,
Phys.\ Rev.\ D {\bf 57}, 2837 (1998)
[arXiv:hep-ph/9710338].

\bibitem{Fujii:1999xn}
H.~Fujii and D.~Kharzeev,
Phys.\ Rev.\ D {\bf 60}, 114039 (1999)
[arXiv:hep-ph/9903495].

\bibitem{Barnes:2003kp}
T.~Barnes and N.~I.~Kochelev,
arXiv:nucl-th/0306026.

\bibitem{Redlich:2000cb}
K.~Redlich, H.~Satz and G.~M.~Zinovjev,
Eur.\ Phys.\ J.\ C {\bf 17}, 461 (2000)
[arXiv:hep-ph/0003079].

\bibitem{Hufner:2000jb}
J.~H\"ufner, Y.~P.~Ivanov, B.~Z.~Kopeliovich and A.~V.~Tarasov,
Phys.\ Rev.\ D {\bf 62}, 094022 (2000)
[arXiv:hep-ph/0007111].

\bibitem{Hufner:1995qe}
J.~H\"ufner and B.~Kopeliovich,
Phys.\ Rev.\ Lett.\  {\bf 76}, 192 (1996)
[arXiv:hep-ph/9504379].

\bibitem{He:1999aj}
Y.~B.~He, J.~H\"ufner and B.~Z.~Kopeliovich,
Phys.\ Lett.\ B {\bf 477}, 93 (2000)
[arXiv:hep-ph/9908243].

\bibitem{Muller:1999ys}
B.~Muller,
Nucl.\ Phys.\ A {\bf 661}, 272 (1999)
[arXiv:nucl-th/9906029].

\bibitem{Kharzeev:1994pz}
D.~Kharzeev and H.~Satz,
Phys.\ Lett.\ B {\bf 334}, 155 (1994)
[arXiv:hep-ph/9405414].

\bibitem{Arleo:2001mp}
F.~Arleo, P.~B.~Gossiaux, T.~Gousset and J.~Aichelin,
Phys.\ Rev.\ D {\bf 65}, 014005 (2002)
[arXiv:hep-ph/0102095].

\bibitem{Oh:2001rm}
Y.~Oh, S.~Kim and S.~H.~Lee,
Phys.\ Rev.\ C {\bf 65}, 067901 (2002)
[arXiv:hep-ph/0111132].

\bibitem{Peskin:1979va}
M.~E.~Peskin,
Nucl.\ Phys.\ B {\bf 156}, 365 (1979).

\bibitem{Bhanot:1979vb}
G.~Bhanot and M.~E.~Peskin,
Nucl.\ Phys.\ B {\bf 156}, 391 (1979).

\bibitem{Kaidalov:hd}
A.~B.~Kaidalov and P.~E.~Volkovitsky,
Phys.\ Rev.\ Lett.\  {\bf 69}, 3155 (1992).

\bibitem{Pes00}
M.~E.~Peskin,
personal communication.

\bibitem{Buchmuller:1980su}
W.~Buchm\"uller and S.~H.~Tye,
Phys.\ Rev.\ D {\bf 24}, 132 (1981).
                                                                                
\bibitem{Blaschke:2000zm}
D.~B.~Blaschke, G.~R.~Burau, M.~A.~Ivanov, Y.~L.~Kalinovsky and P.~C.~Tandy,
arXiv:hep-ph/0002047.

\bibitem{Matinian:1998cb}
S.~G.~Matinian and B.~M\"uller,
Phys.\ Rev.\ C {\bf 58}, 2994 (1998)
[arXiv:nucl-th/9806027].

\bibitem{Lin:1999ad}
Z.~W.~Lin and C.~M.~Ko,
Phys.\ Rev.\ C {\bf 62}, 034903 (2000)
[arXiv:nucl-th/9912046].

\bibitem{Oh:2000qr}
Y.~Oh, T.~Song and S.~H.~Lee,
Phys.\ Rev.\ C {\bf 63}, 034901 (2001)
[arXiv:nucl-th/0010064].

\bibitem{Oh:2002vg}
Y.~Oh, T.~Song, S.~H.~Lee and C.~Y.~Wong,
arXiv:nucl-th/0205065.

\bibitem{Haglin:2003fh}
K.~Haglin and C.~Gale,
arXiv:hep-ph/0305174.

\bibitem{Sibirtsev:2000aw}
A.~Sibirtsev, K.~Tsushima and A.~W.~Thomas,
Phys.\ Rev.\ C {\bf 63}, 044906 (2001)
[arXiv:nucl-th/0005041].

\bibitem{Liu:2001yx}
W.~Liu, C.~M.~Ko and Z.~W.~Lin,
Phys.\ Rev.\ C {\bf 65}, 015203 (2002)
arXiv:nucl-th/0107058.

\bibitem{Haglin:1999xs}
K.~L.~Haglin,
Phys.\ Rev.\ C {\bf 61}, 031902 (2000)
[arXiv:nucl-th/9907034].

\bibitem{Haglin:2000ar}
K.~L.~Haglin and C.~Gale,
Phys.\ Rev.\ C {\bf 63}, 065201 (2001)
[arXiv:nucl-th/0010017].

\bibitem{Martins:1994hd}
K.~Martins, D.~Blaschke and E.~Quack,
Phys.\ Rev.\ C {\bf 51}, 2723 (1995)
[arXiv:hep-ph/9411302].

\bibitem{Wong:1999zb}
C.~Y.~Wong, E.~S.~Swanson and T.~Barnes,
Phys.\ Rev.\ C {\bf 62}, 045201 (2000)
[arXiv:hep-ph/9912431].

\bibitem{Wong:2001td}
C.~Y.~Wong, E.~S.~Swanson and T.~Barnes,
Phys.\ Rev.\ C {\bf 65}, 014903 (2002)
[Erratum-ibid.\ C {\bf 66}, 029901 (2002)]
[arXiv:nucl-th/0106067].

\bibitem{Barnes:2003dg}
T.~Barnes, E.~S.~Swanson, C.~Y.~Wong and X.~M.~Xu,
arXiv:nucl-th/0302052.

\bibitem{Black:2002}
N.~F.~Black,
{\it KN And $J/\psi $N Scattering In The Quark Model}, University of
Tennessee Ph.D. thesis (2002).

\bibitem{Martins:1995nb}
K.~Martins,
Prog.\ Part.\ Nucl.\ Phys.\  {\bf 36}, 409 (1996)
[arXiv:hep-ph/9601314].

\bibitem{Navarra:2001jy}
F.~S.~Navarra, M.~Nielsen, R.~S.~Marques de Carvalho and G.~Krein,
Phys.\ Lett.\ B {\bf 529}, 87 (2002)
[arXiv:nucl-th/0105058].

\bibitem{Duraes:2002ux}
F.~O.~Dur\~aes, H.~Kim, S.~H.~Lee, F.~S.~Navarra and M.~Nielsen,
arXiv:nucl-th/0211092.

\bibitem{Maltman:st}
K.~Maltman and N.~Isgur,
Phys.\ Rev.\ D {\bf 29}, 952 (1984).
                                                                         
\bibitem{Navarra:2001ju}
F.~S.~Navarra, M.~Nielsen and M.~E.~Bracco,
Phys.\ Rev.\ D {\bf 65}, 037502 (2002)
[arXiv:hep-ph/0109188].

\bibitem{Matheus:2002nq}
R.~D.~Matheus, F.~S.~Navarra, M.~Nielsen and R.~Rodrigues da Silva,
Phys.\ Lett.\ B {\bf 541}, 265 (2002)
[arXiv:hep-ph/0206198].

\bibitem{Deandrea:2003pv}
A.~Deandrea, G.~Nardulli and A.~D.~Polosa,
arXiv:hep-ph/0302273.

\bibitem{Barnes:1991em}
T.~Barnes and E.~S.~Swanson,
Phys.\ Rev.\ D {\bf 46}, 131 (1992).

\end{thebibliography}
\end{document}